# Crowd Behaviour during High-Stress Evacuations in an Immersive Virtual Environment

**Mehdi Moussaïd[1*#], Mubbasir Kapadia[2,6#], Tyler Thrash[3#], Robert W. Sumner[2,4], Markus Gross[2,4], Dirk Helbing[5], and Christoph Hölscher[3]**

[1]Center for Adaptive Rationality, Max Planck Institute for Human Development, Berlin, Germany

[2]Disney Research Zurich, Switzerland

[3]ETH Zürich, Chair of Cognitive Science, Zürich, Switzerland

[4]ETH Zürich, Computer Graphics Laboratory, Zürich, Switzerland

[5]ETH Zürich, Computational Social Science, Zürich, Switzerland

[6]Rutgers University, Computer Science Department

* Corresponding author: moussaid@mpib-berlin.mpg.de

[#]MM, MK, and TT contributed equally to this work.

See videos and supplementary material at:

http://rsif.royalsocietypublishing.org/content/13/122/20160414.figures-only

# Abstract

Understanding the collective dynamics of crowd movements during stressful emergency situations is central to reducing the risk of deadly crowd disasters. Yet, their systematic experimental study remains a challenging open problem due to ethical and methodological constraints. In this paper, we demonstrate the viability of shared 3D virtual environments as an experimental platform for conducting crowd experiments with real people. In particular, we show that crowds of real human subjects moving and interacting in an immersive 3D virtual environment exhibit typical patterns of real crowds as observed in real-life crowded situations. These include the manifestation of social conventions and the emergence of self-organized patterns during egress scenarios. High-stress evacuation experiments conducted in this virtual environment reveal movements characterized by mass herding and dangerous overcrowding as they occur in crowd disasters. We describe the behavioral mechanisms at play under such extreme conditions and identify critical zones where overcrowding may occur. Furthermore, we show that herding spontaneously emerges from a density effect without the need to assume an increase of the individual tendency to imitate peers. Our experiments reveal the promise of immersive virtual environments as an ethical, cost-efficient, yet accurate platform for exploring crowd behaviour in high-risk situations with real human subjects.

# Introduction

The dynamics of crowds implies major theoretical and real-world challenges (1–4). As with many other de-centralized social and biological systems (5), the dynamics of crowd movements is driven by nonlinear amplification loops that promote the emergence of large-scale behavioural patterns. Recent progress in modeling and simulation techniques (6, 7), coupled with advances in experimental methods (8, 9) and live monitoring (3, 10–12), has provided unprecedented amounts of theoretical and empirical insights into crowd movements, ranging from the emergence of 'smart' patterns of self-organization to their breakdown when deadly crowd disasters happen (13).

Despite these major advances, one important aspect of crowd behaviour that remains difficult to study is the collective dynamics that takes place under stressful emergency situations (4). Empirical research has reported about several case studies of specific emergency evacuations, such as during the 9/11 attacks (14), the Love parade disaster (15), the Mecca pilgrimage (3, 13), and other fire escape situations (16–18). These works have highlighted prominent features of emergency escapes, such as the preference for familiar exit routes, the feeling of a common social identity within the crowd, and the nature of the fire alarm on people's reaction time. Other studies have demonstrated the contagious aspect of risk perception, suggesting that anxiety may spread from one pedestrian to another during stressful evacuations or that collective underestimation of the danger could lead to critical evacuation delays (19–22).

Yet, fine-grained data analyses are missing to extract the precise mechanisms driving collective behaviours during stressful evacuations. For example, it remains unclear to what extent pushing, overcrowding, and peer imitation can affect the efficiency of egress. The main obstacle to answering these questions is the scarcity of detailed empirical data: Laboratory experiments are not suited for the study of emergency situations due to safety and ethical issues, and real-world observations similar to those described above are rare and difficult to evaluate. Consequently, most research in this domain is conducted by means of computer simulations based on simplified behavioural assumptions (2) or rely on analogies to animal models (23). While computer simulations facilitate the collection of data in a controlled and cost-efficient way, the accuracy of the findings are inherently limited to the extent that the simulations mimic real crowds. Despite promising advances in this area (4, 24), computer simulated agents cannot reliably emulate real human behavior, especially for situations in which empirical data is difficult to obtain in the first place.

To overcome these limitations, we propose a novel approach to study the behaviour of large crowds of real experimental subjects, moving and interacting in shared immersive 3D virtual environments (25, 26). In the last few years, an increasing number of studies have relied on virtual reality devices to investigate the behaviour of pedestrians, for example, by means of head-mounted displays or CAVE systems. Although some limitations were highlighted, such as a gender bias in handling the navigation controls (27), simple navigation tasks and route choice experiments were successfully conducted in virtual environments (28–30). Virtual worlds have also been used to study features of emergency evacuations, but social interactions among pedestrians were absent (31) or limited to a single subject facing a group of simulated agents (32–35).

One crucial aspect of crowd dynamics lies in the social interactions that take place between individuals. These interactions create feedback loops and amplification effects and give rise to self-organized macroscopic patterns. It is therefore important to observe groups of participants moving and interacting simultaneously in the same environment. Notably, one study has managed to study crowd evacuation with groups of real people navigating simultaneously in the virtual world provided by the game Second Life (36), but the constraints imposed by the game structure made it difficult to keep a good control on all experimental variables.

This new technique is in concert with the recent development of computational methods in the social sciences that employ artificial environments to study the dynamics of large social systems (37), such as cultural markets (38), social networks (39), and collective problem-solving (40). Here, we extend this experimental principle to crowd behaviour by allowing a large number of participants to navigate freely in an immersive virtual space and interact with one another in real-time. This experimental technique enables the systematic study of crowd dynamics under extreme conditions, with complete control of experimental variables and without the prohibitive safety and ethical concerns of real-world experiments.

In the present experiment, 36 experimental subjects participated simultaneously. Each participant sat in front of a computer screen and had a first-person view of the surrounding virtual environment, including the other participants (see **Figure 1**). Subjects navigated freely in the environment by using the computer mouse and the keyboard (see Materials and Methods, and **figure S1**). We first assessed the validity of the method by replicating a series of previously conducted real-world crowd experiments using our virtual world platform (Studies 1 and 2). At both the micro and macro levels of observation, the virtual environment turns out to be a good proxy for real-life dynamics. Then, we explored the dynamics of high-stress evacuations in a series of experiments for which participants have to evacuate a building on fire under strong time pressure and heavy monetary penalization in case of failure (Study 3). We observed realistic panic movements and analysed emerging patterns of overcrowding and collective route choice. Our study demonstrates the promise of immersive multi-user virtual environments for the study of crowd dynamics, which opens a wide variety of research and applications.

# Results

**Method validation.** In Study 1, we replicated a real-life experiment in which pairs of pedestrians are instructed to avoid each other in a narrow corridor (41) (**Figure S3**). The avoidance behaviors in the virtual environment conformed to real-life observations in terms of the shape of the trajectories and the choice of the passing side (**Figure 2A**). Interestingly, we observed a marked side-preference during avoidance maneuvers in the virtual environment **(Figure 2B)**. In more than 95% of our replications, experimental subjects chose to avoid each other on the right-hand side. A two-proportion Z-test was used to compare the proportion of replications in which participants passed each other on the right side to a chance value of 50%, Z=17.03, p<.001. This finding indicates that participants in the virtual corridor were following an existing social convention during avoidance maneuvers. In the real-life experiment, 81% of the subjects avoided towards the right-hand side, but these proportions cannot be directly compared because the participants were drawn from different populations (i.e., in France for the real-life experiment and Switzerland for the virtual experiment). The main conclusion is that participants exhibited a marked side preference in the virtual corridor, suggesting that virtual worlds also capture some social aspects of pedestrian behaviour.

Study 2 tested the reliability of the virtual environment for reproducing *collective* crowd patterns. We studied 36 subjects performing a series of evacuation tasks, emulating the experimental design from Ref. (42). Participants were immersed in a large virtual room and instructed to evacuate through a bottleneck of varying width, ranging from 60cm to 150cm (**Figures 1 and S4**). Consistent with real-life findings, the outflow of pedestrians increased linearly with the bottleneck width (**Figure 3**). When compared to a larger body of real-life datasets, the outflow of participants seemed to be smaller in the virtual environment. This discrepancy can be due to a multitude of micro-navigation factors, such as differences in walking speed, acceleration, or the shoulder movements between real and virtual environments. Although not identical, the observed trends in the virtual world are reasonably similar to the real-life dynamics to consider virtual environments as proxies for real-life dynamics.

**Implementation of emergency evacuations.** In Study 3, we performed a series of emergency egress experiments for low-stress ($C_0$) and high-stress ($C_1$) conditions (**Figure 4A**). The environment consisted of a complex building with four possible exit locations $E_1$, $E_2$, $E_3$, and $E_4$ through which participants were instructed to escape (**Figures 4B and S5**). For each replication, the functional exit door was placed at a randomly chosen exit location, whereas the other three exit locations are blocked. Participants were unaware of the location of the functional exit door, except for a certain proportion $k$ of informed individuals who could see an arrow in the top of the screen that indicated the direction of the safe exit (43). Participants knew that some group members may have been informed of the correct exit but cannot recognize them, thus mimicking the social uncertainty of real-life egress.

Stress is implemented by manipulating three experimental factors: (*i*) *Time pressure*: Participants had to escape the building within 50 seconds for $C_1$. No time limit was imposed for $C_0$. (*ii*) *Reward system*: Throughout the experiment, participants could collect points that were converted into monetary bonuses at the end of the session (see Material and Methods). In condition $C_0$, participants were *rewarded* 50 points upon escaping the building. In condition $C_1$, however, participants were *penalized* 100 points if they did not manage to escape in time, with no bonus for a successful escape. The reward system was therefore switched from the gain domain to the loss domain under high-stress (44). For both $C_0$ and $C_1$, participants were additionally penalized 1 point for colliding with another participant or obstacle. (*iii*) *Environmental factors*: A series of stress-inducing elements in the environment were implemented in $C_1$ but not in $C_0$. These elements included lower luminosity, red blinking lights, and fires at the blocked exit locations.

**Dynamics of emergency escape.** We observed noticeable behavioural differences between the two conditions. In the absence of stress, participants tended to keep reasonably safe distances from their neighbors in order to avoid the collision penalty. Consequently, body contacts hardly occurred during low-stress evacuations **(Figures 4 and S6),** as in similar real-life situations. In contrast, a high frequency of body contacts occurred in the high-stress condition, despite the application of the same collision penalty. Therefore, participants appeared ready to lose a considerable amount of points due to body collisions — and to impose the same penalty to their neighbors — to maximize the likelihood of escaping on time. On average, participants lost nearly the same amount of points due to body collisions (26 points per replication, SD=13) and due to failures to escape (36 points per replication, SD=38).

The density levels also reflected relative crowdedness. It remained lower than 2 person/m$^2$ in $C_0$, which is typically observed in everyday congested zones. Under high stress, however, the density level reached values up to 5 p/m$^2$, which violated all safety standards and was close to the critical threshold of crowd turbulence (13). The most dangerous zones with the highest density levels were (*i*) areas in which a decision needed to be made, (*ii*) areas surrounding the exit where bottlenecks occurred and caused congestion, and (*iii*) dead ends where the flow of people returning after exploring a wrong option encountered with the flow of those moving in the opposite direction (**Fig. S7 and S8**). This overcrowding pattern was not only due to the reduction of interpersonal distance, but also due to the fact that most people *decided* to go in the same direction. We characterized the herding level *H(t)* at each time *t* by measuring *H(t)=p_{maj}(t)-p_{min}(t)*, where *p_{maj}(t)* represents the proportion of uninformed individuals who chose the branch where the majority of individuals converged at the end of the replication (analogously, *p_{min}(t)* stands for the minority proportion). In $C_0$, $p_{maj}$ and $p_{min}$ tended to increase at the same rate corresponding to an *H*-value close to 0 (**Figure 5A**). After approximately 45 seconds, the flow of people who made an incorrect first decision reached the other branch, which was reflected by the subsequent gradual increase of *H(t)*. In $C_1$, however, the great majority of people chose the same branch at the beginning, and the herding level H approached 1 after a short time. In order to evaluate this pattern statistically, we fit linear regression models to each trial and compared the slopes of the best fit lines for $C_0$ to the slopes of the best fit lines for $C_1$. As predicted, the slopes from $C_1$ were significantly greater than the slopes from $C_0$, t(14) = 6.65, p <, .001, d = 0.12, even after accounting for differences with respect to the length of each trial, t(14) = 7.52, p < .001, d = 0.02. What are the behavioural mechanisms underlying the emergence of this herding pattern? We hypothesize that, under time and monetary pressure, subjects would increase their tendency to follow their neighbors as suggested in an early model of crowd panics (2), which would give rise to the observed herding pattern under high-stress.

We tested this assumption by considering the response function $f(S)$, which describes the individual probability to choose one branch or the other as a function of the social signal *S* produced by the crowd at the moment of a decision. In our experiment, the social signal $S(t)$ is the movement of the crowd in the main corridor at the time of decision *t*, formally defined as:

$$S(t) = \sum_i^{N'} v_i^x(t).$$

Here, $v_i^x(t)$ is the horizontal component of participant *i*'s velocity indicating whether participant *i* was moving towards the right or the left side of the floor plan, and *N'* is the subset of participants who were present in the main horizontal corridor at time *t*. The empirically determined response function $f(S)$ has a typical S-shape (**Figure 5B**), indicating that individuals make use of social information when deciding where to go (45). Surprisingly, however, the response functions measured under low-stress and high-stress conditions were quite similar, which was at odds with our first intuition. In order to evaluate this similarity statistically, we compared the correspondence between the response functions for $C_0$ and $C_1$ to the correspondence between randomly generated datasets. For values of $f(S)$ that were missing for either $C_0$ or $C_1$, we randomly generated numbers between 0 and 1 from continuous, uniform distributions. One thousand replacement values were generated in this way for each missing value. For each set of original data with some proportion of randomly generated values, we then calculated the correlation between $C_0$ and $C_1$. We also randomly generated 1000 pairs of whole datasets and calculated the correlation between $C_0$ and $C_1$ in a similar way in order to produce a null distribution of correlation coefficients. An independent-samples t-test determined that the set of correlation coefficients derived from the original data was significantly greater than the correlation coefficients derived from random datasets, $t(1998)$ = 92.31, $p$ < .001, $d$ = 1.04. Note that, although this approach is relatively unsophisticated, it is also conservative compared to other approaches that replace missing values using the distribution of the original data (e.g., multiple imputation; (46)).

Further analysis revealed that individuals were exposed to much stronger social signals under high-stress than low-stress situations due to increased local density levels, as shown by the distributions of $|S|$ in **Figure 5C**. While the values of $|S|$ were lower than 5 in 75% of the decisions made in low-stress conditions, the distribution was positively skewed under high-stress conditions and included values up to 35. Therefore, the *same* response function $f(S)$ held in both conditions but applied to higher values of $S$ under high-stress than under low-stress conditions. Put simply, pedestrians had a higher probability of following their neighbors when stress was high, simply because the neighboring individuals were *more numerous* due to the increased density level. Herding, therefore, resulted from the crowdedness and not from a change in the individual tendency to imitate neighbors.

# Discussion

The collective dynamics that takes place during stressful emergency evacuations is probably the least understood aspect of crowd behavior, despite being crucial for crowd safety. In this work, we have proposed to observe crowds of *real* human subjects moving and interacting in *virtual* environments. Our approach offers important advantages and opens numerous research perspectives. First, it resolves safety and ethical issues and enables the systematic exploration of crowd behaviour under high-stress conditions with real human participants. Second, it is flexible and enables the exploration of crowd behaviours in potentially any virtual place without restrictions in terms of environment topology or size. Third, it allows for a rich variety of measured variables with high accuracy, including participants' field of view, and can be combined with eye-tracking or physiological measurement devices. Fourth, it permits the accurate control and manipulation of experimental variables such as light level, walking speed, and body sizes.

We validated our experimental platform with respect to its ability to reliably replicate the dynamics of real crowds and demonstrate its potential to conduct previously infeasible studies such as the study of crowd behaviour in high-stress evacuations. However, our platform may benefit from further improvement. As the results of Study 2 suggest, the virtual environment will require further calibration work. While the bottleneck experiment in the virtual environment reproduces real-life data reasonably well, we have noted that the flow values were considerably lower in the virtual setting than in the real world. This difference could be due to a variety of differences at the micro-navigation level, such as dissimilarities in walking speeds or time delays when a keyboard key is pressed or released. Future work would therefore need to calibrate the control interface in order to produce more realistic crowd movements.

In the current state of development, organizing experiments with a larger number of participants (i.e., greater than the maximum capacity of the computer laboratory, 36) remains as difficult as for real-life experiments. In either case, experimental subjects need to be physically present in an experimental room, which involve other logistic challenges when the number of participants is large. As a consequence, between-group replications may be scarce. In our data, for example, we could not completely rule out group-specific biases (e.g, habituation effects), although none were detected (see **Figure S9**). This issue could possibly be addressed by extending the experimental platform to a Web version for which participants would not need to be physically present in the laboratory (38, 39, 47). Future work will therefore focus on extending our laboratory-based experimental approach to web-based experiments, facilitating between-group replications and extending the number of simultaneous participants.

Our results leave open interesting questions that could be addressed in future studies. As past research has shown, social identification among individuals tends to promote inter-individual cooperation and enhance the efficiency of emergency evacuations (3, 33). The reward system that we have implemented and the separation of people into cubicles could have encouraged participants to behave in a more competitive manner, which could explain the observed number of collisions. In future works, social identification level could be manipulated experimentally in the virtual environment to address this issue.

According to current research on collective movements, interaction networks based on sensory information (e.g., vision) are crucial to understand emerging movement patterns (1, 10, 48). One important direction of future work will therefore focus on establishing networks of visual contacts to determine precisely how visual cues propagate from person to person and how this information impacts herding behaviours (49). This could be inferred from the position of individuals in the environment or by means of eye trackers.

In conclusion, the use of immersive multi-user virtual environments promises to be a powerful tool that can push the boundaries of crowd research in new and exciting directions, enabling new applications for urban planners and architects. Such applications include evaluating the quality of service and evacuation plans of building designs in virtual reality.

# Methods

**Experimental software**. The experimental software was developed using the Unity3D game engine (Unity Technologies), ADAPT (50) for animating the virtual characters, and SmartFoxServer for the networking procedures. The platform immerses participants in a visually realistic virtual environment in which all users can freely navigate and see the other participants in real time. Subjects had a first-person view of the environment and could navigate by means of a keyboard and mouse. Navigation included three degrees of freedom: forward/backward translations, left/right translations, and left/right rotations (**Figure S1 and Video S1**). The control interface was tested in a previous study (51) and yielded the best navigation performance as compared to two other control solutions (keyboard-only and joystick) with respect to real human walking trajectories. For simplicity, we assume homogeneous virtual characters (height: 1.8m, shoulder width: 0.25 m, maximum forward walking speed: 1.3 m/s, backwards and lateral moving speed: 0.6 m/s). A circular collision check with a diameter equal to the shoulder width was implemented to ensure that virtual pedestrians do not overlap in crowded situations.

**Experimental design**. Two experimental sessions took place in June and December 2014. For each session, 36 experimental subjects were hired and invited to the laboratory. They received between 20 and 50 CHF for their participation depending on performance. Data was collected in the ETH Decision Science Lab (DeSciL) which independently approved the experimental procedures according to its human subjects regulations. Informed consent was obtained from all participants based on DeSciL lab requirements. Participants were seated in a room containing 36 cubicles, each containing a desktop computer. They could not see the screen of the other participants and were not allowed to communicate with each other during the experiment. Subjects were instructed to wear headphones for the duration of the experiment. Each experimental session started with a training phase of approximately 40 minutes, during which all participants learned how to navigate in the virtual environment. In this phase, subjects had to complete a step-by-step tutorial designed to review all possible movements (**Figure S2**). During the first experimental session, we conducted Studies 1 and 2, while Study 3 was conducted during the second session. Throughout each experiment, subjects earned points that were converted to monetary compensation at the end of the session. In all experiments, participants were penalized 1 point every time they collided with another participant or obstacle. Participants initially started with 1000 points in the second session to compensate for the expected losses from the high-stress experiment.

**Study 1** is the replication of a real-life experiment conducted previously (41) in which pairs of participants moving in opposite directions had to avoid each other in a narrow corridor. The 36 subjects were randomly grouped in pairs and placed at each end of a straight corridor (length = 8m; width = 1.8m; **Figure S3**). Each participant was instructed to reach the other end of the corridor without colliding. Any collision with the corridor walls or the other participant resulted in a penalty of 1 point. At the end of each replication, new pairs of participants were randomly assigned. The 18 pairs of subjects performed the experiment simultaneously in 18 independent virtual corridors. We collected 561 replications of this experimental condition in approximately 10 minutes, which illustrates the flexibility of our experimental platform.

**Study 2** is a replication of a real-life evacuation experiment conducted previously (42). Participants were initially located in a large room (width = 10m; length = 4m) and instructed to walk through a bottleneck after the starting signal to a finish line located 10 meters after the bottleneck (**Figure S4**). The bottleneck width varied from 0.6m to 1.5m. We performed 14 replications in total, two for each bottleneck width. Two replications were later discarded because some participants deliberately blocked the outflow by standing in front of the bottleneck door. Participants received a bonus of 100 points after reaching the finish line and had no incentives for completing the task faster than others. The different bottleneck widths appeared in a random order.

**Study 3** was divided into a first block of 10 replications for the low-stress condition and a second block of 12 replications for the high-stress condition. Participants did not see the map of the environment, but they were allowed to explore it freely during a preliminary training session. In the low-stress condition, subjects were instructed to find the exit door of a complex building (**Figure S5**). No time limit was imposed to find the exit door, and subjects were awarded 50 points at the end of each replication. The high-stress condition was the same except for the 3 following stress-inducing factors: (*i*) A time limit of 50 seconds was imposed. The time limit was calibrated such that participants had enough time to explore one exit but not enough to explore a second one if the first option was not correct. (*ii*) Subjects who did not manage to escape within the time limit received a penalty of 100 points. Those who were successful did not receive any additional bonus. (*iii*) A set of stress-inducing elements were added to the environment including red blinking lights, lower luminosity, fire blocking the wrong exit doors, and the sounds of an alarm. In each replication of the low- and high-stress conditions, a certain proportion *k* of subjects were informed about the location of the exit. Informed participants could see an arrow on the top of their screen pointing towards the exit. All subjects knew that some of them could be informed but did not know how many and could not recognize informed individuals. We varied the proportion of informed subjects in k: 0%, 10%, 33%, and 100%. The purpose of this manipulation was to give participants the feeling that some of their neighbors might know the location of the exit, which mimics the uncertainty of real-life evacuations. The proportion of informed individuals k, as well as the location of the exit door were randomized between trials.

## Acknowledgements
We thank C. Wilhem, F. Thaler, H. Abdelrahman, and H. Zhao for helpful assistance during the software development. We are also grateful to G. De Polavieja, and J. Gouello for insightful discussions.

## Additional Information

**Author contributions**: MM, MK, TT, RS, MG, DH and CH designed research. MM, MK, and TT performed research, analyzed data, and wrote the paper. MM, MK, and TT contributed equally to this work. All authors reviewed the manuscript.
**Competing interests**: The authors declare no competing interests.
**Funding:** Funding was received from D.H.'s ERC Advanced Investigator Grant "Momentum" (Grant No. 324247). This research was also supported by a grant from the German Research Foundation (DFG) as part of the priority program on *New Frameworks of Rationality* (SPP 1516) given to Ralph Hertwig and Thorsten Pachur (HE 2768/7-2). The funders had no role in study design, data collection and analysis, decision to publish, or preparation of the manuscript.

# References

1. Moussaïd M, Helbing D, Theraulaz G (2011) How simple rules determine pedestrian behavior and crowd disasters. *Proceedings of the National Academy of Sciences* 108(17):6884–6888.

2. Helbing D, Farkas I, Vicsek T (2000) Simulating dynamical features of escape panic. *Nature* 407(6803):487–490.

3. Alnabulsi H, Drury J (2014) Social identification moderates the effect of crowd density on safety at the Hajj. *Proceedings of the National Academy of Sciences* 111(25):9091–9096.

4. Schadschneider A, et al. (2009) Evacuation Dynamics: Empirical Results, Modeling and Applications. *Encyclopedia of Complexity and Systems Science* (Springer New York), pp 3142–3176.

5. Couzin I, Krause J (2003) Self-organization and collective behavior in vertebrates. *Adv Stud Behav* 32:1–75.

6. Raupp Musse S, Thalmann D (2001) Hierarchical model for real time simulation of virtual human crowds. *IEEE Trans Vis Comput Graph* 7(2):152–164.

7. Duives DC, Daamen W, Hoogendoorn SP (2013) State-of-the-art crowd motion simulation models. *Transp Res Part C: Emerg Technol* 37:193–209.

8. Hoogendoorn S, Daamen W (2005) Pedestrian Behavior at Bottlenecks. *Transportation Science* 39(2):147–159.

9. Seyfried A, et al. (2009) New Insights into Pedestrian Flow Through Bottlenecks. *Transportation Science* 43(3):395–406.

10. Gallup A, et al. (2012) Visual attention and the acquisition of information in human crowds. *Proceedings of the National Academy of Sciences* 109(19):7245–7250.

11. Wirz M, et al. (2013) Probing crowd density through smartphones in city-scale mass gatherings. *EPJ Data Science* 2(1):1–24.

12. Wagoum AUK, Seyfried A, Holl S (2012) Modeling The Dynamic Route Choice Of Pedestrians To Assess The Criticality Of Building Evacuation. *Adv Complex Syst* 15(07):1250029–1–1250029–22.

13. Helbing D, Johansson A, Al-Abideen H (2007) The Dynamics of crowd disasters: an empirical study. *Physical Review E* 75(4):46109.

14. Johnson CW Lessons from the Evacuation of the World Trade Centre, September 11th 2001 for the Development of Computer-Based Simulations. Available at: http://www.dcs.gla.ac.uk/~johnson/papers/9_11.PDF.

15. Helbing D, Mukerji P (2012) Crowd disasters as systemic failures: analysis of the Love Parade disaster. *EPJ Data Science* 1(1):1–40.

16. Proulx G (2001) Occupant behaviour and evacuation. *Proceedings of the 9th International Fire Protection Symposium* (archive.nrc-cnrc.gc.ca), pp 219–232.

17. Proulx G, Sime JD (1991) To prevent'panic'in an underground emergency: why not tell people the truth? *Fire Safety Science* 3:843–852.

18. Sime JD (1983) Affiliative behaviour during escape to building exits. *J Environ Psychol* 3(1):21–41.

19. Nilsson D, Johansson A (2009) Social influence during the initial phase of a fire evacuation—Analysis of evacuation experiments in a cinema theatre. *Fire Saf J* 44(1):71–79.

20. Moussaïd M, Brighton H, Gaissmaier W (2015) The amplification of risk in experimental diffusion chains. *Proc Natl Acad Sci U S A* 112(18):5631–5636.

21. Faria J, Krause S, Krause J (2010) Collective behavior in road crossing pedestrians: the role of social information. *Behav Ecol* 21(6):1236–1242.

22. Scherer C, Cho H (2003) A Social Network Contagion Theory of Risk Perception. *Risk Anal* 23(2):261–267.

23. Saloma C, Perez GJ, Tapang G, Lim M, Saloma P (2003) Self-organized queuing and scale-free behavior in real escape panic. *Proceedings of the National Academy of Science* 100(21):11947.

24. Zheng X, Zhong T, Liu M (2009) Modeling crowd evacuation of a building based on seven methodological approaches. *Build Environ* 44(3):437–445.

25. Warren W, Kay B, Zosh W, Duchon A, Sahuc S (2001) Optic flow is used to control human walking. *Nat Neurosci* 4(2):213–216.

26. Fajen B, Warren W (2003) Behavioral dynamics of steering, obstacle avoidance, and route selection. *J Exp Psychol Hum Percept Perform* 29(2):343–362.

27. Lucas K, Sherry JL (2004) Sex Differences in Video Game Play:: A Communication-Based Explanation. *Communic Res* 31(5):499–523.

28. Kretz T, et al. (2011) Calibrating dynamic pedestrian route choice with an Extended Range Telepresence System. *Computer Vision Workshops (ICCV Workshops), 2011 IEEE International Conference on*, pp 166–172.

29. Schrom-Feiertag H, Schinko C, Settgast V, Seer S (2014) Evaluation of Guidance Systems in Public Infrastructures Using Eye Tracking in an Immersive Virtual Environment. *ET4S@ GIScience*, pp 62–66.

30. Vilar E, et al. (2013) Are Emergency Egress Signs Strong Enough to Overlap the Influence of the Environmental Variables? *Design, User Experience, and Usability. User Experience in Novel Technological Environments*, Lecture Notes in Computer Science. (Springer Berlin Heidelberg), pp 205–214.

31. Kinateder M, et al. (2015) The effect of dangerous goods transporters on hazard

perception and evacuation behavior--A virtual reality experiment on tunnel emergencies. *Fire Saf J* 78:24–30.

32. Kinateder M, et al. (2014) Social influence on route choice in a virtual reality tunnel fire. *Transp Res Part F Traffic Psychol Behav* 26, Part A(0):116–125.

33. Drury J, et al. (2009) Cooperation versus competition in a mass emergency evacuation: a new laboratory simulation and a new theoretical model. *Behav Res Methods* 41(3):957–970.

34. Bode NWF, Kemloh Wagoum AU, Codling EA (2014) Human responses to multiple sources of directional information in virtual crowd evacuations. *J R Soc Interface* 11(91):20130904.

35. Bode N, Codling E (2013) Human exit route choice in virtual crowd evacuations. *Anim Behav* 86(2):347–358.

36. Normoyle A, Drake J, Safonova A (2012) Egress Online: Towards leveraging massively, multiplayer environments for evacuation studies. Available at: http://citeseerx.ist.psu.edu/viewdoc/download?doi=10.1.1.278.5452&rep=rep1&type=pdf.

37. Ball P (2007) Social science goes virtual. *Nature* 448:647–648.

38. Salganik M, Dodds P, Watts D (2006) Experimental study of inequality and unpredictability in an artificial cultural market. *Science* 311(5762):854–856.

39. Centola D (2010) The Spread of Behavior in an Online Social Network Experiment. *Science* 329(5996):1194–1197.

40. Mason W, Watts D (2012) Collaborative learning in networks. *Proceedings of the National Academy of Sciences* 109(3):764–769.

41. Moussaïd M, et al. (2009) Experimental study of the behavioural mechanisms underlying self-organization in human crowds. *Proceedings of the Royal Society B: Biological Sciences* 276(1668):2755–2762.

42. Kretz T, Grünebohm A, Schreckenberg M (2006) Experimental study of pedestrian flow through a bottleneck. *J Stat Mech* 2006(10):P10014.

43. Couzin I, Krause J, Franks N, Levin S (2005) Effective leadership and decision-making in animal groups on the move. *Nature* 433(7025):513–516.

44. Tversky A, Kahneman D (1974) Judgment under Uncertainty: Heuristics and Biases. *Science* 185(4157):1124–1131.

45. Strandburg-Peshkin A, Farine DR, Couzin ID, Crofoot MC (2015) Shared decision-making drives collective movement in wild baboons. *Science* 348(6241):1358–1361.

46. Schafer JL, Graham JW (2002) Missing data: our view of the state of the art. *Psychol Methods* 7(2):147–177.

47. Lazer D, et al. (2009) Computational social science. *Science* 323(5915):721–723.

48. Rosenthal SB, Twomey CR, Hartnett AT, Wu HS, Couzin ID (2015) Revealing the hidden networks of interaction in mobile animal groups allows prediction of complex behavioral contagion. *Proc Natl Acad Sci U S A* 112(15):4690–4695.

49. Strandburg-Peshkin A, et al. (2013) Visual sensory networks and effective information transfer in animal groups. *Curr Biol* 23(17):R709–11.

50. Shoulson A, Marshak N, Kapadia M, Badler NI (2014) ADAPT: The Agent Development and Prototyping Testbed. *IEEE Trans Vis Comput Graph* 20(7):1035–1047.

51. Thrash T, et al. (2015) Evaluation of Control Interfaces for Desktop Virtual Environments. *Presence: Teleoperators and Virtual Environments* 24(4):322–334.

52. Liddle J, et al. (2009) An Experimental Study of Pedestrian Congestions: Influence of Bottleneck Width and Length. *Traffic and Granular Flow 2009* doi:citeulike-article-id:6203264.

53. Daamen W, Hoogendoorn S (2010) Capacity of doors during evacuation conditions. *Procedia Engineering* 3(0):53–66.

# Figures

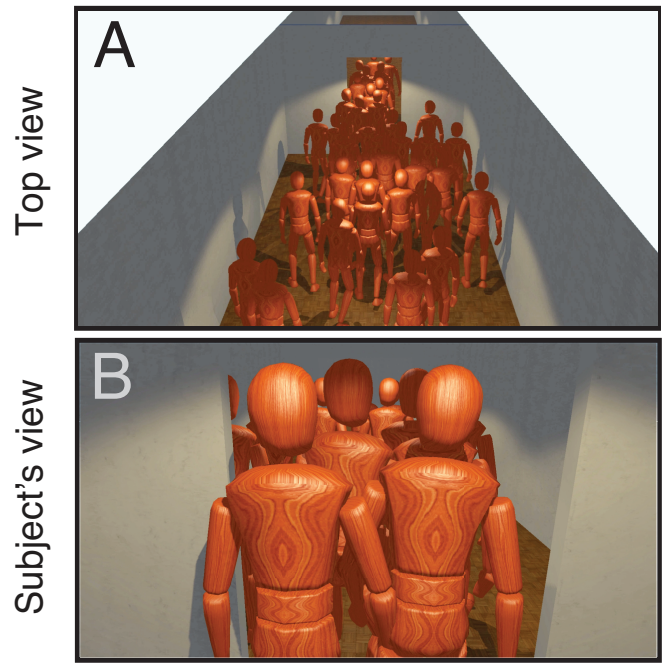

**Figure 1**: **Illustration of the virtual environment**. (A) Top-down view of a crowd of 36 participants passing through a bottleneck during a simple evacuation situation. Each pedestrian in this snapshot was controlled by a real experimental participant who can navigate freely in the environment. (B) First-person view of the same situation as seen by one participant located in the middle of the crowd.

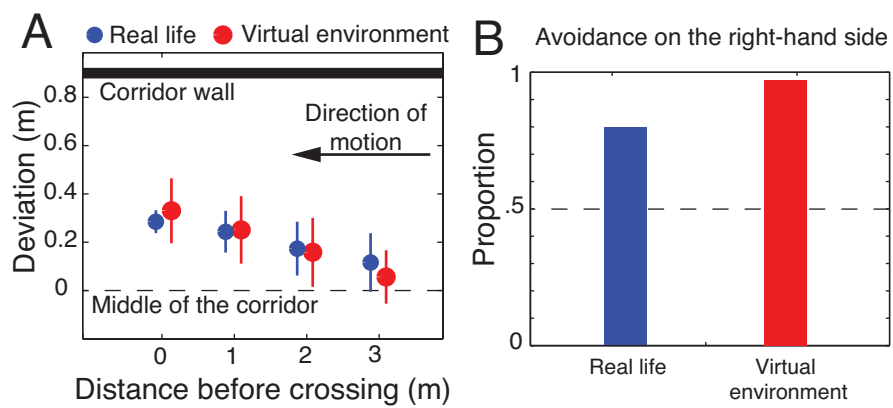

**Figure 2**: **Comparison of virtual and real results during simple avoidance maneuvers** (41, 42). (A) Average lateral deviation of the walking trajectories during a simple avoidance task for which two participants moving in opposite directions avoid each other in a narrow corridor (real-life experiment in blue, N=144; virtual environment in red, N=561). Error bars indicate the standard deviation of the mean. (B) Proportion of participants avoiding each other on the right-hand sides during the same experiments.

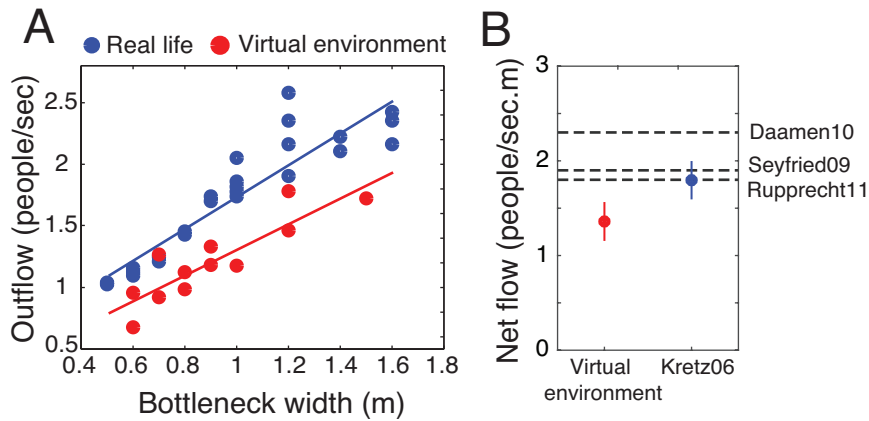

**Figure 3: Flow through a bottleneck in real and virtual environment.** (A) Flow of people through bottlenecks of varying width, measured during a group evacuation experiment in our virtual environment (red dots), replicating real-life experiments (blue dots). Lines of best fit are $f(x) = 1.29x + 0.45$ and $f(x) = 1.04x + 0.27$ for the real-life and virtual environments, respectively. (B) Net flow per unit of door width in the virtual environment (in red) and for study (42) (in blue). Error bars indicate the standard deviation of the mean. The three black dashed lines show the average values reported in three other real-life studies (9, 52, 53).

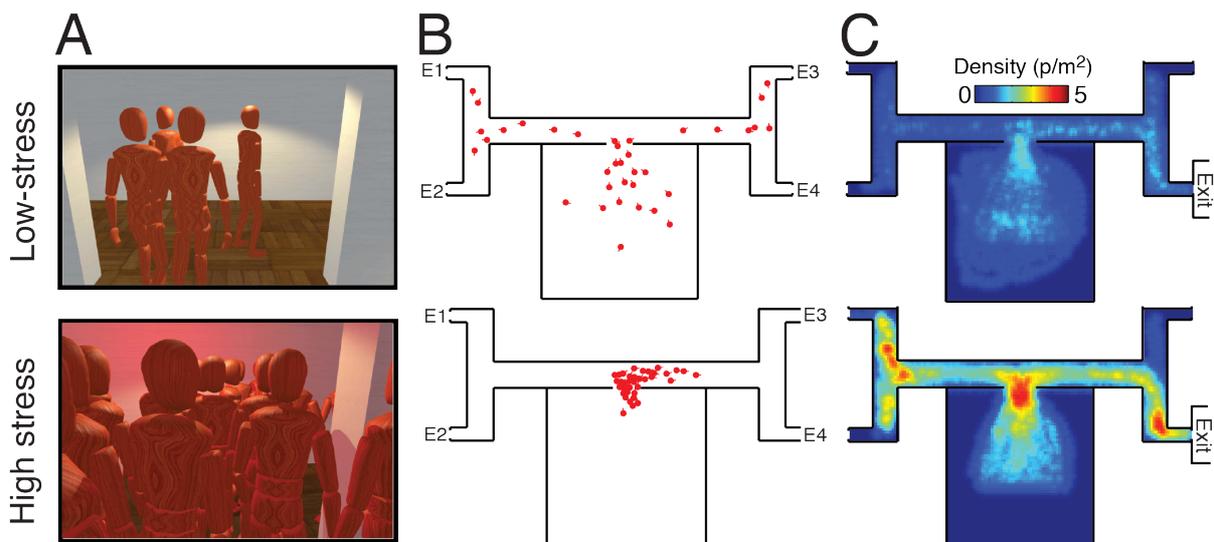

**Figure 4**: **Comparison between low-stress and high-stress escape experiments.** (A) Illustrative snapshots of the environment as seen by one participant approaching the decision zone in the low-stress (top) and high-stress (bottom) conditions. In addition to time pressure and the risk of losing money, the high-stress environment is characterized by stress-inducing factors such as lower luminosity and red blinking lights. (B) Representative top-down views of the participants' positions in both conditions. Under low stress, participants kept a certain distance from each other and tended to explore both branches of the main corridor. Under high stress, participants were densely packed and herded in the same branch. For each replication, the free exit door was randomly placed at one location among $E_1$, $E_2$, $E_3$, and $E_4$. In both examples illustrated here, the exit door was located at position $E_2$. (C) Maximum density levels measured all over the environment, averaged across all replications ($N_{low\_stress}$=10 and $N_{High-stress}$=12). Density levels hardly approach 2 p/m$^2$ in low-stress conditions, but reached up to 5 p/m$^2$ under high-stress — a very high value at which physical injuries might occur.

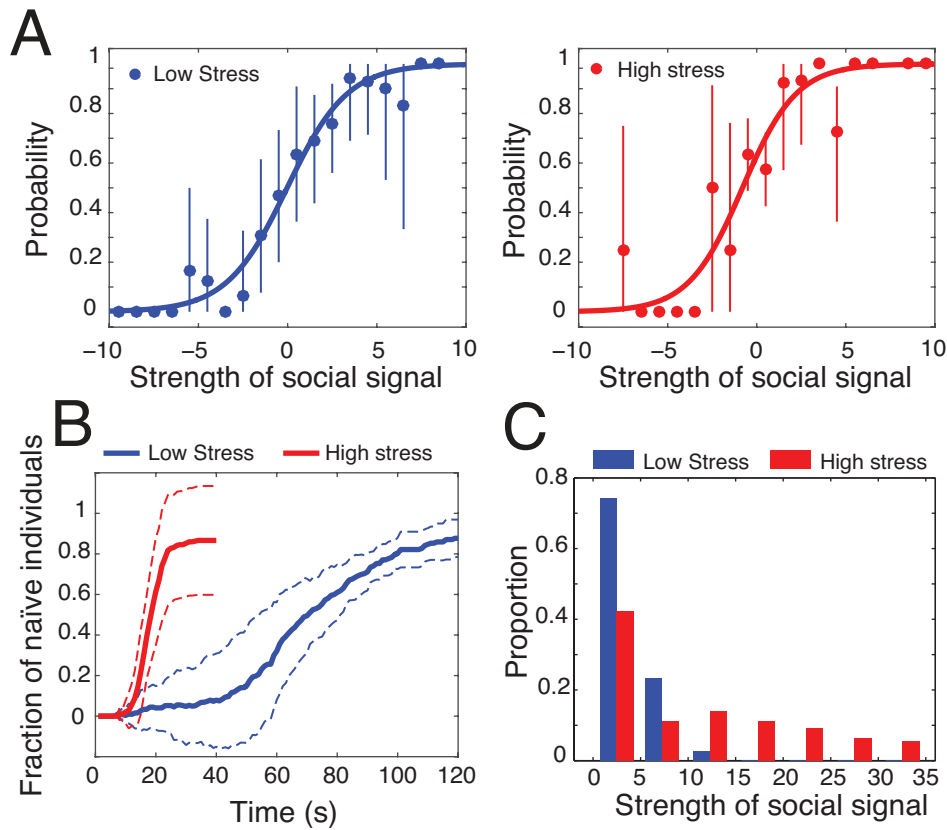

**Figure 5**: **Herding dynamics.** (A) Individual probabilities to choose the right-hand branch when arriving in the decision zone as a function of the social signal produced by the crowd at that moment. A positive signal indicates crowd movements directed towards the right side (a negative one, respectively, towards the left side). The left and right figures correspond to the low-stress and high stress conditions, respectively. The response function was almost identical in both conditions, indicating that the observed herding patterns do not result from a change in the herding tendency but instead from the crowdedness. The fitted curves were obtained by minimization of the squared distance to the data points using the equation $1/(1 + e^{aS+b})$, resulting in $a = -0.59$ and $b = 0.03$ under no stress and $a = -0.66$ and $b = 0.80$ under high-stress. (B) Average herding level *H(t)* indicating the fraction of uninformed individuals who chose the same branch as the majority of individuals, under low-stress (blue) and high-stress (red) conditions. The dashed lines represent the standard deviation of the average. Herding is stronger under high stress than under low stress (also illustrated in **Figure 4B**), despite a similar individual response function shown in (A). (C) The distribution of the social signal strength shows that the social signal is weaker under low stress (blue) than under high stress (red).